\documentclass{PoS}
\def\be{\begin{equation}}
\def\ee{\end{equation}}
\title{Large N transition in the 2D SU(N)xSU(N) 
nonlinear sigma model. }

\ShortTitle{Large N transition in the 2D SU(N)xSU(N) 
nonlinear sigma model. }

\author{\speaker{Rajamani Narayanan}%
\\
        Florida International University, Department of Physics, 
Miami, FL 33199, USA\\
        E-mail: \email{rajamani.narayanan@fiu.edu}}

\author{Herbert Neuberger\\
        Department of Physics and Astronomy, Rutgers University,
Piscataway, NJ 08855, USA\\
        E-mail: \email{neuberg@physics.rutgers.edu}}

\author{Ettore Vicari\\
        Dipartimento di Fisica, Universit\'a di Pisa and INFN,
I-56127 Pisa, Italy\\
        E-mail: \email{vicari@df.unipi.it}}

\abstract{We consider the characteristic polynomial associated
with the smoothed two point function in 
two dimensional large $N$ principal
chiral model. We 
numerically show that it undergoes a transition at a critical
distance of the order of the correlation length. The transition
is in the same universality class as two dimensional large $N$
QCD.
}

\FullConference{The XXVI International Symposium on Lattice Field Theory\\
		 July 14-19 2008\\
		 Williamsburg, Virginia, USA}

\begin{document}

\section{Two dimensional SU(N) X SU(N) principal chiral model}

The two dimensional SU(N) X SU(N) principal chiral model is
similar to four dimensional SU(N) gauge theory in many
respects\cite{Rossi:1996hs}.
The continuum action is given by
\be
S=\frac{N}{T}\int d^2 x Tr \partial_\mu g(x) \partial_\mu g^\dagger (x)
\ee
where $g(x) \in$ SU(N).
The global symmetry group SU(N)${}_L\times$
SU(N)${}_R$ reduces down to a single SU(N)
``diagonal subgroup'' if we make
a translation breaking ``gauge choice'', $g(0)=1$.
This model is asymptotically free and there are $N-1$
particle states with masses
\be
M_R=M\frac{\sin(\frac{R\pi}{N})}{\sin(\frac{\pi}{N})},~~~1\le R\le N-1.
\ee
The states corresponding to the $R$-th mass are a multiplet 
transforming as an $R$ component antisymmetric tensor of the
diagonal symmetry group.

The two point function
$W=g(0) g^\dagger(x)$ plays the role of Wilson loop
with the separation $x$ playing the role of area.
We expect the behavior to be perturbative for small
$x$. On the other hand, non-perturbative effects become
important for large $x$.

One expects 
\be
G_R(x)=\langle \chi_R (g(0) g^\dagger (x) ) 
\rangle \sim C_R {N\choose R} e^{-M_R |x|}
\ee
where $\chi_R$ is the trace in the $R$-antisymmetric representation.
Comparison with the heat-kernel representation
of the characteristic polynomial associated with the
Wilson loop operator in two dimensional large $N$ 
QCD~\cite{Narayanan:2007dv}
suggests the following connections:
\begin{itemize}
\item The two point correlator, $W(d)=g(0) g^\dagger (d)$,
is analogous to the Wilson loop operator.
\item $M|x|$ is analogous to the dimensionless area, $t$. 
\end{itemize}
Based on this analogy, we hypothesize~\cite{Narayanan:2008he}
 that the
characteristic polynomial, $\det\left(z-  g(0) g^\dagger (d)\right)$,
will undergo a transition at some value $d_c$. The
universal behavior at this transition will be in the same
universality class as two dimensional large $N$ QCD.

\section{Setting the scale}

Numerical measurement of the correlation length using
the lattice action 
\be
S_L = -2 N b \sum_{x,\mu} \Re Tr [ g(x) g^\dagger(x+\mu)]
\ee
and
\be
\xi_G^2 =\frac{1}{4} \frac{\sum_x x^2 G_1(x)}{\sum_x G_1(x)}
\ee
yields the following continuum result~\cite{Rossi:1993zc}:
\be
M\xi_G =0.991(1)
\ee

We use $\xi_G$ to set the scale and it is
well described by
\be
\xi_G = 0.991~ \left [ 
\frac{e^{\frac{2-\pi}{4}} }{16\pi} \right ] ~\sqrt{E} ~\exp\left (\frac{\pi}{E}\right )
\ee
in the range $11 \le \xi_G \le 20$ with
\be
E=1-\frac{1}{N}\Re\langle Tr [g(0) g^\dagger({\hat 1} )] \rangle
=\frac{1}{8b}+\frac{1}{256 b^2} +\frac{0.000545}{b^3}- \frac{0.00095}{b^4}+ \frac{0.00043}{b^5}
\ee
The above equations will be used to find a $b$ for a given $\xi$.

\section{Smeared SU(N) matrices}
Well defined operators are obtained using smeared matrices.
We start with $g(x)\equiv g_0 (x)$ and
one smearing step takes us from $g_t(x)$ to
$g_{t+1}(x)$ using the following procedure.
Define $Z_{t+1} (x)$ by:
\be
Z_{t+1} (x)=\sum_{\pm\mu} [g^\dagger_t (x) g_t (x+\mu)-1 ].
\ee
Construct anti-hermitian traceless $SU(N)$ matrices $A_{t+1} (x)$ 
\be
A_{t+1}(x)=Z_{t+1}(x)-Z^\dagger_{t+1}(x) -\frac{1}{N} \rm{Tr} (Z_{t+1}(x)-Z^\dagger_{t+1}(x))\equiv-A^\dagger_{t+1}(x).
\ee
Set
\be
L_{t+1}(x)=\exp[{f A_{t+1}(x)}].
\ee
$g_{t+1}(x)$ is defined in terms of $L_{t+1} (x)$ by:
\be
g_{t+1} (x)= g_t (x) L_{t+1} (x).
\ee
This procedure is iterated till we reach $g_n(x)$ and
the smearing parameter is defined by $\tau= nf$.
For a fixed $\xi_G$, the parameter $\tau$ is fixed
such that $\tau/\xi_G^2$ remains unchanged.
We set $n=30$ in our numerical simulations and this
was found sufficiently large to eliminate a dependence on
the two factors, $f$ and $n$, individually.

\section{Numerical details}

We need $L/\xi_G > 7$ to minimize finite volume effects.
We worked in the range $11 \le \xi_G \le 20$
and therefore we chose $L=150$.
We used a combination of Metropolis and over-relaxation
at each site $x$ for our updates. The full SU(N) group was
explored.
200-250 passes of the whole lattices were sufficient
to thermalize starting from $g(x)\equiv 1$.
50 passes per step were enough to equilibrate if $\xi_G$ was
increased in steps of $1$.

The test of the universality hypothesis proceeds in the same
manner as for the three dimensional large $N$ gauge theory.
We defined the characteristic polynomial, $F(y,d)$, as
\be
F(y,d)=\langle\det(e^{y/2}+e^{-y/2} W(d))\rangle
\ee
We perform a Taylor expansion,
\be
F(y,d,N)
=C_0(d,N) +C_2(d,N) y^2 +C_4(d,N) y^4+\dots
\ee
since $F(y,d)$ is an even function of $y$.
It is useful to define
\be
\Omega(d,N) = \frac{ C_0(d,N) C_4(d,N)}{C_2^2(d,N)}
\ee
which resembles a Binder cumulant.

\begin{figure}
\vskip 1cm
\centerline{\includegraphics[width=0.8\textwidth]{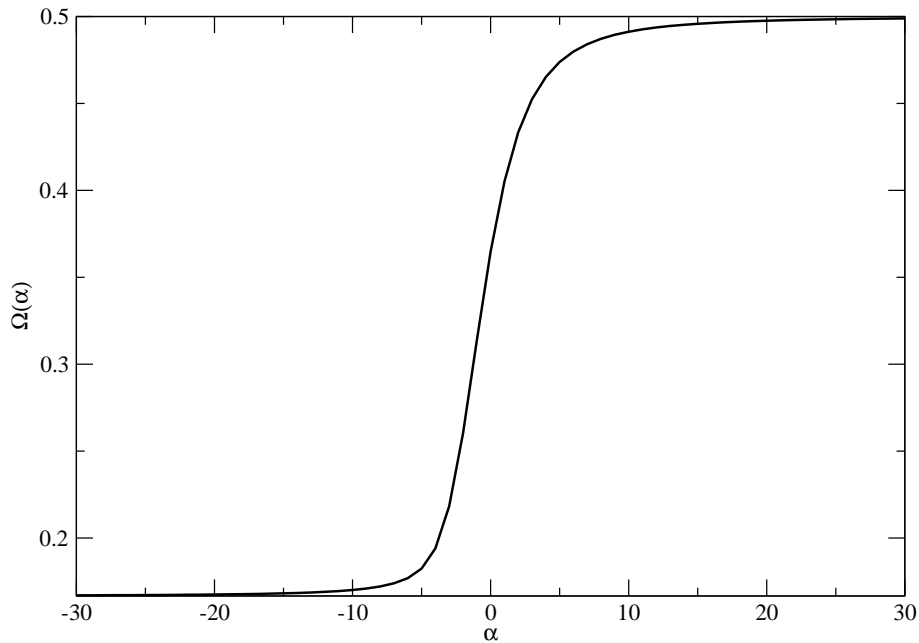}}
\caption{Behavior of $\Omega$ as a function of $\alpha$
in the scaling region.\label{omegas}}
\end{figure}

As $N\to\infty$,
$\Omega(d,\infty)$ is a step function with
$\Omega=\frac{1}{6}$ for short distances
$d < d_c $ and $\Omega=\frac{1}{2}$ for long distances,
$d > d_c $. 
Zooming in on the step function as $N\to\infty$ in the
vicinity of $d=d_c$  using the scaling variable
$\alpha \propto \sqrt{N} (d-d_c)$, we obtain Fig.~\ref{omegas}.

\begin{figure}
\vskip 1cm
\centerline{
\includegraphics[width=0.8\textwidth]{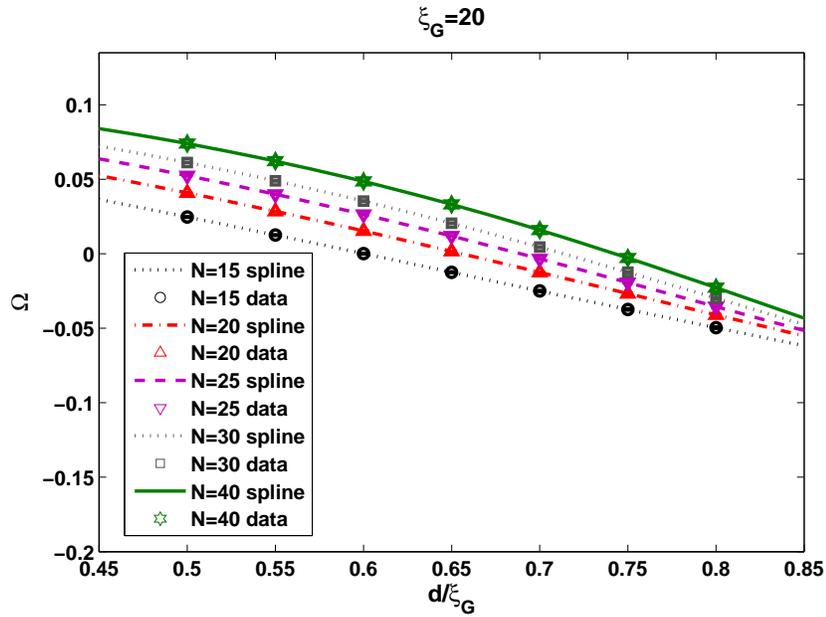}
}
\caption{Plot of $\Omega(d)$ after the subtraction of
$\Omega(\alpha=0)=0.364739936$ as a function of $d/\xi_G$.
\label{firstfig}}
\end{figure}

\begin{figure}
\vskip 1cm
\centerline{
\includegraphics[width=0.8\textwidth]{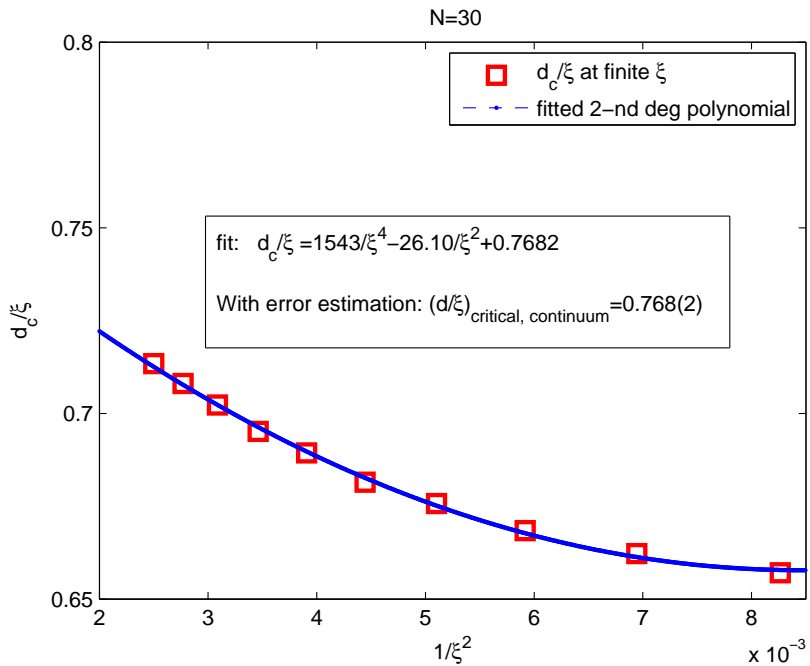}
}
\caption{Extrapolation to continuum of $d_c/\xi$ for $N=30$.
\label{secondfig}}
\end{figure}

\begin{figure}
\vskip 1cm
\centerline{
\includegraphics[width=0.8\textwidth]{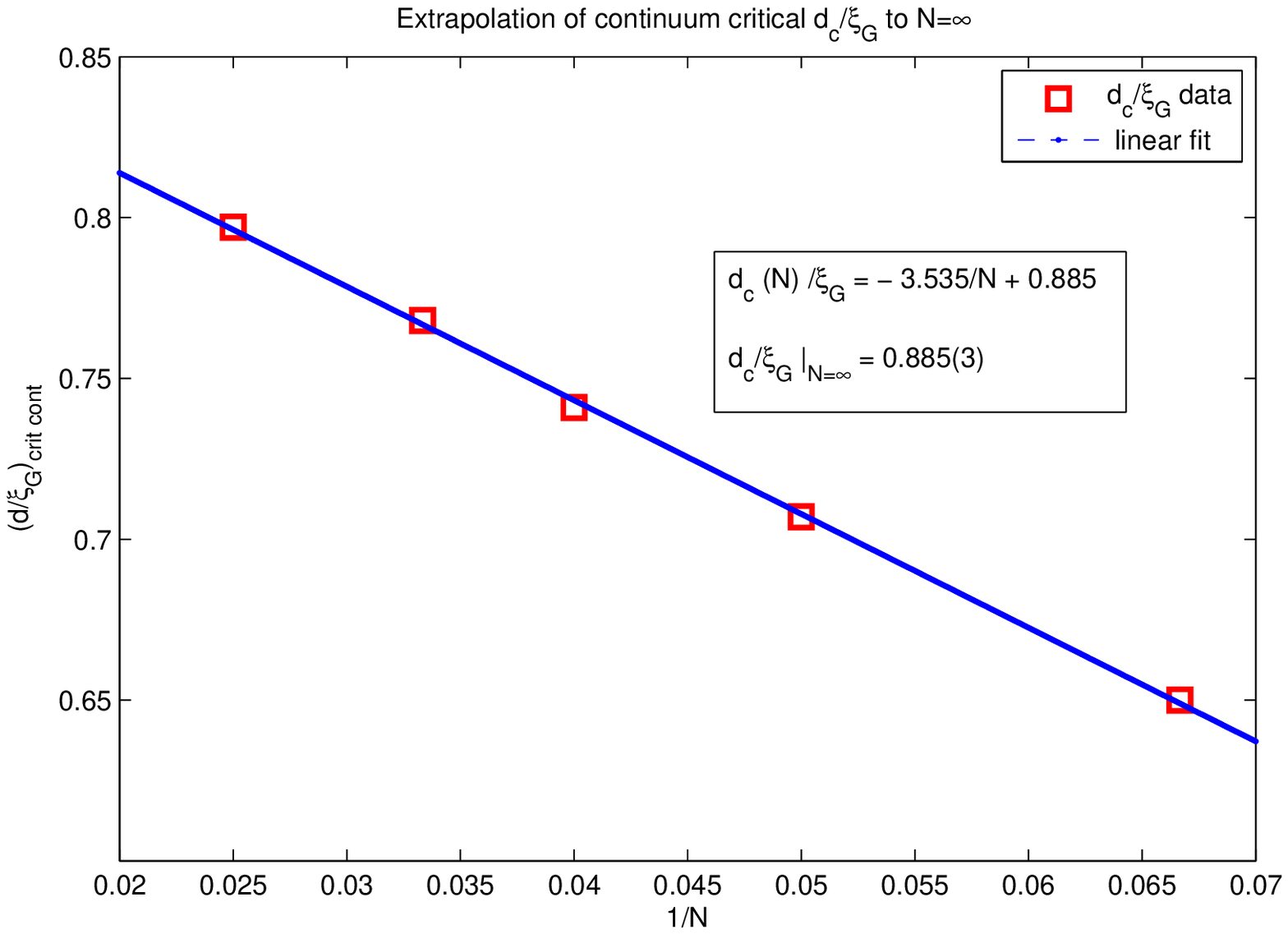}
}
\caption{Extrapolation of the continuum $d_c/\xi$ to
infinite $N$.\label{thirdfig}}
\end{figure}

We use $\Omega(\alpha=0)=0.364739936$ to obtain the
critical size $d_c$ in the following manner.
Given an $N$ and a $\xi$, we find the  $d_c$ that
makes the Binder cumulant $\Omega(d_c,N)=0.364739936$
as shown in Fig.~\ref{firstfig}.
We look at $d_c$ as a function of $\xi$ for a given $N$.
This gives us the continuum value of $d_c/\xi$ for that $N$.
This extrapolation is shown in Fig.~\ref{secondfig} for
$N=30$.
We then take the large $N$ limit as shown in
Fig.~\ref{thirdfig} and it gives us
\be
\frac{d_c}{\xi_G} |_{N=\infty}=0.885(3)
\ee

\begin{figure}
\vskip 1cm
\centerline{
\includegraphics[width=0.8\textwidth]{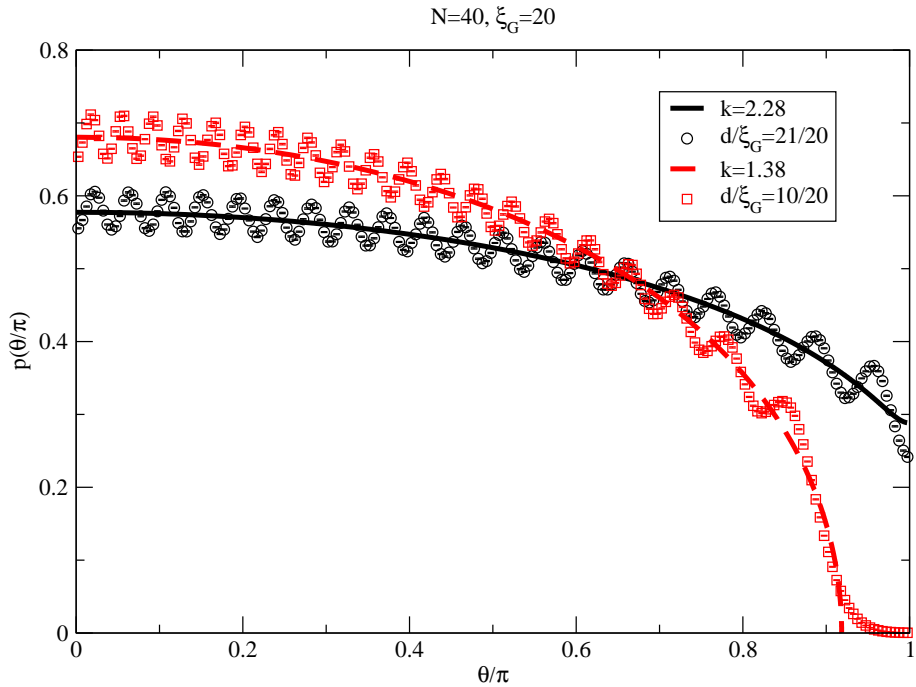}
}
\caption{Examples of eigenvalue distribution for one small
and one large distance.\label{doa}}
\end{figure}

\begin{figure}
\vskip 1cm
\centerline{
\includegraphics[width=0.8\textwidth]{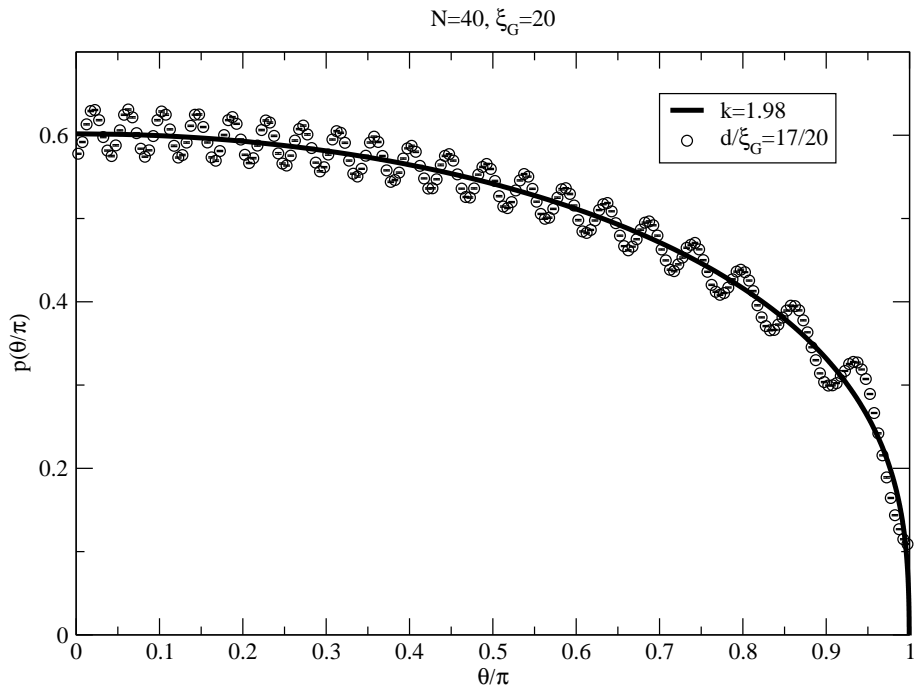}
}
\caption{An example of an almost critical eigenvalue
distribution.\label{dob}}
\end{figure}

Further substantiation of the universal behavior can
be given by comparing the eigenvalues distribution in
the model to the 
Durhuus-Olesen eigenvalue distributions in two
dimensional QCD. This is shown for one example
each on either side of
the critical point
in Fig.~\ref{doa} 
and very close to the critical point
in Fig.~\ref{dob}. We use $2k=t$ to match with
the notation in~\cite{Durhuus:1980nb}.

\acknowledgments

R.N. acknowledge partial support by the NSF under grant number
PHY-055375 at Florida International University.  
H. N. acknowledges partial support by the DOE, grant \#
DE-FG02-01ER41165, and the SAS of Rutgers University.

\end{document}